# Microtubule shuttles on kinesin-coated glass micro-wire tracks

Kyongwan Kim,[a] Andrew L. Liao,[a,b] Aurélien Sikora,[a] Daniel Oliveira,[a] Hikaru Nakazawa,[c] Mitsuo Umetsu,[a,c] Izumi Kumagai,[c] Tadafumi Adschiri,[a] Wonmuk Hwang[b,d,e] and Winfried Teizer*[a,b,f]

[a] WPI-Advanced Institute for Materials Research (AIMR), Tohoku University, Sendai 980-8577, Japan.
[b] Materials Science and Engineering, Texas A&M University, College Station, TX 77843-3003, USA.
[c] Department of Biomolecular Engineering, Graduate School of Engineering, Tohoku University, Sendai 980-8579, Japan.
[d] Department of Biomedical Engineering, Texas A&M University, College Station, TX 77843-3120, USA.
[e] School of Computational Sciences, Korea Institute for Advanced Study, Seoul 130-722, Korea.
[f] Department of Physics and Astronomy, Texas A&M University, College Station, TX 77843-4242, USA.
*E-mail:teizer@physics.tamu.edu, Phone: 1-979-845-7730, Fax: 1-979-845-2590

*Abstract*

Gliding of microtubule filaments on surfaces coated with the motor protein kinesin has potential applications for nano-scale devices. The ability to guide the gliding direction in three dimensions allows the fabrication of tracks of arbitrary geometry in space. Here, we achieve this by using kinesin-coated glass wires of micrometer diameter range. Unlike previous methods in which the guiding tracks are fixed on flat two-dimensional surfaces, the flexibility of glass wires in shape and size facilitates building in-vitro devices that have deformable tracks.

## 1. Introduction

Microtubules are cytoskeletal filaments formed by noncovalent self-assembly of tubulin dimers as building blocks. A tubulin dimer (46 Å × 80 Å × 65 Å) consists of α- and β-tubulin (Desai and Mitchison, 1997; Nogales et al., 1999). A microtubule typically consists of 13 protofilaments self-assembled into a tubule 25 nm in diameter (Chretien et al., 1995). The motor protein kinesin 'walks' on the microtubule to carry out many vital functions in the cell, including intracellular transport and cell division (McIntosh, 2012; Hirokawa, 2009; Vale 2003). Kinesin uses adenosine triphosphate (ATP) as a fuel and its 8 nm step size corresponds to the size of the tubulin pitch forming the microtubule (Vale and Milligan, 2000; Hwang and Lang 2009). Due to their physiological importance, pharmaceutical targeting of microtubules and kinesin is under intensive study for diseases such as Alzheimer's disease and cancer (Zhou and Giannakakou, 2005; Craddock et al., 2012; Rath and Kozielski, 2012). Moreover, by virtue of in-vitro formation of microtubules (Weisenberg, 1972) and successful reconstruction of the kinesin/microtubule-based motility (Vale et al., 1985; Vale et al., 1985), this bio-motile system has been utilized as a model system for the study of pattern formation driven by motor proteins (Nedelec, 1997; Schaller et al., 2010; Sumino et al., 2012). Unlike in traditional approaches where one passively observes self-organization of a dynamic system, motor proteins allow controlled experimental design, which is a significant benefit for the study of phase transitions in systems of self-driven components (Vicsek et al., 1995).

In addition, the robust stepping process (8 nm stepsize for kinesin on microtubules (Coy et al., 1999)), activation at the single molecule level (Schnitzer and Block, 1997), as well as the intrinsic size of the bio-structures, are opening a new paradigm for nano/bio-mechanical device development. Although the landing of kinesin molecules on the microtubule track is a random event, the bound kinesin always moves in a specific direction defined by the polarity of the microtubule. This peculiarity in kinesin motility is understood as a result of the broken symmetry in the binding between the kinesin motor head and the tubulin lattice (Grant et al., 2011) as well as the unidirectional force generation upon conformational changes in the kinesin motor head (Hwang, 2008, Lakkaraju and Hwang, 2011). The asymmetric, processive and autonomous movement of kinesin on a microtubule meets the requirements for a 'molecular walker' (von Delius and Leigh, 2011).

Besides motility of individual kinesins on the microtubule (Sikora et al., 2012), a cooperative mode of transport can be realized in-vitro, where microtubules glide on kinesin-coated surfaces (Cohn et al., 1989; Howard et al., 1989). Collective operation of the motor domains holding one microtubule results in its translocation, allowing its use as a potential transporter (Brunner et al., 2007). This cooperative mode has a great advantage in that the distance of travel of microtubules is several orders of magnitude longer compared to the run lengths of an individual kinesin (Agarwal and Hess, 2010; Muthukrishnan et al., 2009). However, for a practical application of the bio-motile system randomness in the initial gliding direction of microtubules still needs to be overcome. Namely, in order to apply gliding microtubules driven by molecular motors to a nanotransport device for sorting, separation, or assembly of materials, it is essential to guide microtubules along predetermined pathways (van den Heuvel and Dekker, 2007). Various approaches to design guiding tracks have been investigated, including topographically patterned tracks (Hess et al., 2001; Hiratsuka et al., 2001), chemical patterning of the kinesin attachment sites (Clemmens et al., 2003), a combination of physical and chemical approaches (Cheng et al., 2005), and enclosed microfluidic channels as means for directional confinement (van den Heuvel et al., 2006; Yokokawa et al., 2006).

In this work, we use a glass wire as a new type of guiding track for the kinesin-powered gliding of microtubules. A glass surface is a traditional choice for bio-motile systems as a majority of the bio-motility research has been demonstrated within a thin glass chamber, the so-called flow cell. The central idea of this work is the reduction in dimensionality of motion from a two-dimensional surface to a quasi one-dimensional wire. Glass fibers have been widely used in optical communication as a track to guide photons. Methods to fabricate glass micro/nano wires and to shape them are well established (Tong et al., 2003). Our approach of using glass wires for guiding microtubule motility thus has the following advantages: (1) Unlike previous methods in which guiding tracks are rigidly defined within a two-dimensional surface, the glass-wire track performs its function essentially in three-dimensional space. (2) Whereas previous guiding tracks are mostly enclosing types in terms of geometrical configuration, our wire track is an exposed type. Thus, tracks are not independently isolated but can be coupled to each other so that proteins on a track can interact with those on another track. (3) Glass wires are flexible at the micro/nano scale and each track can be manipulated individually in principle. These features can be useful for making reconfigurable devices for molecular transport.

**2. Methods**

2.1. Glass wire

Glass wires were fabricated by a customized method (Tong et al., 2006). The experimental set-up and a schematic diagram of the procedure are shown in Figures 1a and 1b, respectively. First, a sapphire rod (Optostar Ltd., Diameter: 0.97 mm ± 0.02 mm, Length: 100 mm) was heated by a butane torch. Once the temperature of the rod was sufficient to melt glass, we brought its tip in contact with a piece of coverslip in order to generate a droplet of melted glass held by the tip. We then brought another sapphire rod (Optostar Ltd., Diameter: 0.45 mm ± 0.02 mm, Length: 100 mm) in contact with the glass droplet while the temperature of the first rod was kept constant. While reducing the temperature by removing the flame slightly from the original position the thinner (second) sapphire rod was drawn until a silky glass wire, which formed between the two sapphire rods, disconnected. Because the melting point of sapphire (~2000 °C) is higher than the highest temperature achievable with a butane torch (~1400 °C) the sapphire rods are almost permanently reusable.

Commercially available streptavidin coated quantum dots (Invitrogen Ltd. Qdot 655-streptavidin conjugates) exhibit a strong affinity for glass surfaces in an aqueous environment. The fabricated glass wires were immersed in a quantum dot solution (diluted in PEM buffer: 80 mM PIPES, 1 mM EGTA, 1 mM $MgCl_2$) and then transferred into a droplet of PEM buffer on a glass coverslip for fluorescence imaging. Exemplary quantum dot-decorated glass wires with lateral dimensions ~2.6 μm (determined at the intensity of 7500 in Figure 1e) and ~0.6 μm (determined at the intensity of 3100 in Figure 1f) are shown in Figures 1c and 1d, respectively. The two ends of the wire in Figure 1d are out-of-focus because the wire transits the focal plane (Supplementary Movie, SM01).

2.2. Flow cell

Flow cells including glass wires were assembled by a standard method using double-sided tape (3M 665) as a spacer (Noel et al., 2009). First, we defined a flow channel on a glass slide using double-sided tape. Next, we placed the prepared glass wire on the tape in a way that it crossed the channel. Generally, the taper-drawn glass wire gets gradually thinner towards the open end. The thinner portion of the wire, which we put inside the flow channel, is very flexible and hardly visible to the naked eyes. Therefore, the glass wire was manipulated as it was still attached to the sapphire rod. After putting the wire on the tape, it was cut off the sapphire rod using a razor blade. Finally, the channel was capped with a glass coverslip (Figure 2).

2.3. Microtubule and kinesin

Microtubules were polymerized from commercially available porcine tubulin following a protocol described in previous work (Maloney et al., 2011). For fluorescence imaging, we used a 7:3 mixture of bare and rhodamine-labeled tubulin (Cytoskeleton, Inc.). Tubulin stock solution (5 mg/mL) was prepared in PEM buffer with 1 mM of GTP and glycerol at 6 % (v/v) and stored in 1 μl aliquots (flash-frozen by immersing the vials in liquid nitrogen) at -80 °C. For motility assay experiments microtubules were polymerized by incubating a fresh 1 μl tubulin aliquot at 37 °C for 30 min. The polymerized microtubules were stabilized and diluted by adding 199 μl pre-warmed (at 37 °C for 5 min) taxol solution (10 μM in PEM).

Kinesin motor proteins were expressed in *Escherichia coli* (*E-coli*) by following a standard protocol (Coy et al., 1999; Oliveira et al., 2012). The expression plasmid of kinesin, a truncated *Drosophila* kinesin heavy chain (400 residues, starting from MSAEREIPAE~ ending in ~EDLMEASTPN), followed by the biotin attachment domain and the hexa histidine-tag, was constructed in a pRA vector. The plasmid contains double-resistance against ampicilin and chloramphenicol. The kinesin plasmid inserted *E-coli* BL21 (DE3) cells underwent colonization on a LB agar plate. Isolated colonies were transferred into starter LB medium cultures (12 tubes of 3 mL) and shaken at 28 °C overnight (TAITEK Bio-shaker BR-40LF, 150 rpm). After the initial amplification the starter cultureswere transferred into a 2×YT broth (total volume 2 L divided in 12 flasks). All of the growth media contained two antibiotics (ampicilin 0.1 mg/mL, chloramphenicol 34 μg/mL). Adding 0.1 mM IPTG and 50 μM biotin, the culture flasks were shaken at 15 °C for 24 hrs (TAITEK Bio-shaker BR-300LF, 130 rpm) for kinesin expression and biotin attachment. *E-coli* pellets obtained from centrifugation (TOMY NA-18HS rotor, 4 °C, 6,300 rpm, 10 min) of the culture media were resuspended in the lysis buffer (50 mM Tris, 200 mM NaCl, 500 mM imidazole, 100 μM MgATP, 5 mM BME, pH8-HCl balance), lysed by sonication (three times, 30 sec each) and cooled for 2 min on ice. Cellular debris after lysing was removed by centrifugation (TOMY MX305, 4 °C, 10,000 g, 30 min). The supernatant was transferred into new tubes, centrifuged once more, decanted, and kept at 4 °C. Purification of the histidine-tagged kinesin proteins was performed with a 2 ml Ni-sepharose column. First, the column was equalized with 20 mL of equilibrium buffer (50 mM Tris, 200 mM NaCl, 100 μM MgATP, 5 mM BME, pH8-HCl balance). Next, the supernatant was passed through the column. Then the column was washed with 20 mL of washing buffer (50 mM Tris, 200 mM NaCl, 40 mM imidazole, 100 μM MgATP, 5 mM BME, pH8-HCl balance). Finally, the kinesin was eluted with 10 mL of the elution buffer (50 mM Tris, 200 mM NaCl, 500 mM imidazole, 100 μM MgATP, 5 mM BME, pH8-HCl balance). The purified kinesin solution was separated by gel-filtration chromatography (Superdex 200 column). SDS-PAGE was carried out in order to choose fractions including kinesin proteins. The selected fractions were collected and dialyzed for buffer exchange (storage buffer: 50 mM imidazole, 100 mM NaCl, 1 mM $MgCl_2$, 2 mM EGTA, 0.1 mM EDTA, 5 % (w/v) sucrose, 5 mM BME, 1 mM MgATP, pH7-HCl balance). After filter-centrifugation (TOMY MX305, at 4 °C, 4000 g, Filter: Millipore, MW=10 kDa) to concentrate the resulting solution, we carried out UV-spectroscopy measurements to estimate the kinesin concentration. The final kinesin solution was stored in 5~10 µl aliquots (flash-frozen by immersing the vials in liquid nitrogen) at -80 °C.

2.4. Motility assay

For microtubule gliding experiments, kinesin was thawed and diluted in PEM buffer with 10 mM MgATP. Taxol-stabilized microtubules (see the microtubule and kinesin section) were diluted 10-fold in PEM buffer including 10 mM MgATP and 10 μM Taxol. For a stable and long lasting fluorescence observation, the rhodamine fluorophores were stabilized by adding BME (β-mercaptoethanol, 0.5 % (v/v)) and a typical anti-fade system (20 μg/mL glucose oxidase, 8 μg/mL catalase, 20 mM glucose) into the PEM buffer (Maloney et al., 2011). First, the glass surfaces within a flow cell including a glass wire were exposed to kinesin molecules by introducing 20 μL of kinesin solution (~ 4 μM kinesin, 10 mM ATP in PEM buffer). A pipette and a piece of filter paper were used to inject the fluid to one open side of the flow cell and to absorb it from the other side, respectively. The flow cell was incubated at room temperature for 5min, followed by introduction of a 20-μL microtubule solution. Microtubules were imaged using a fluorescence microscope (Olympus, IX71) equipped with a CCD camera (Hamamatsu, ImagEM) and rhodamine filter set (Omega Optical, Inc., XF204). Image analyses were carried out using Metamorph and ImageJ. Activity of molecular motors fixed on a surface is very sensitive to interaction with the surface. For the optimal functionality and effective consumption of the kinesin proteins in microtubule gliding assays, casein is typically employed to pretreat glass surfaces (Maloney et al., 2011). However, inclusion of casein in our experimental procedure often suppressed binding of microtubules to the kinesin-treated surface, possibly due to a non-optimal casein solution. For the present work, we coated the glass surface densely with kinesin as an alternative way of surface passivation (Howard et al., 1989; Liu et al., 2011). Assuming that all kinesin molecules in solution within the flow cell attach to the glass surfaces, the kinesin surface density from the 4 μM kinesin solution is estimated to be 108,000 molecules per 1 $μm^2$. This is well beyond the saturation limit determined by the size of a single kinesin motor head because it corresponds to one kinesin molecule in an area of ~3 nm × 3 nm which is smaller than the size of a kinesin motor head (~8 nm in diameter) (Kozielski et al., 1997).

## 3. Results and discussion

Figure 3 shows fluorescence microscope images of microtubules in a flow cell including a glass wire, captured at three different focal planes of the same area (The schematic diagram below the image shows the focal plane. See Supplementary Movie, SM02). On the coverslip, only microtubules bound to the surface are clearly recognized (Figure 3a). When the focal plane is near the center of the glass wire, most microtubules including those on the glass wire are out of focus, but the outline of the wire can be identified (Figure 3b). The white guidelines used in Figures 3a and 3c were obtained from Figure 3b. Figure 3c shows microtubules attached to the top of the glass wire. With proper adjustment of the position of focal plane and contrast, the microtubules on the glass wire can visually be separated from the background objects for clear observation.

The gliding motion of microtubules confined by the glass wire (Supplementary Movie, SM03) is visualized by eight images in Figure 4, where the focal plane was set around the glass wire. The white dashed lines again indicate the glass wire identified in Figure 4b. The focal plane was formed at the coverslip surface for the first frame (Figure 4a) and the last frame (Figure 4h) where the change in the distribution of microtubules is due to the gliding motion on the kinesin-coated coverslip. In Figure 4b the focal plane was set around the glass wire for its identification. The focal plane was elevated to the upper portion of the glass wire for the other fiveframes (Figure 4c-4g) to detect microtubules gliding on top of the wire.

In order to characterize the general aspect of the motility, we observed microtubules gliding on a region of the coverslip (area: ~50 μm × 50 μm) away from the glass wire (Supplementary Movie, SM12). Note that all observations in this work were made in a single flow cell. Therefore, there was no significant difference in environmental conditions between observations. One end of each microtubule was tracked for three minutes. The selected ends were marked every 30 sec. Displacement was defined by the length of a straight line connecting the initial mark and the final mark. The average speed was estimated by dividing the linear displacement by the 30-s time interval between frames. The direction of motion was measured relative to the 12 o'clock direction of the image (angle increases counterclockwise). Tracking continued until the microtubule tip moved out of the field of view, overlapped with other microtubules, or detached from the glass surface. Similarly, we analyzed the motion of microtubules on ten different portions of the glass wire whose diameter varied from ~2.6 μm to ~6.5 μm (See Supplemental Movies, SM02 ~SM11). In this case, the direction of motion was measured relative to the vector pointing upwards parallel to the glass wire (arrow in Figure 4b; the angle increases again counterclockwise.)

Figure 5a compares velocities of microtubules moving on the glass coverslip (blue squares) and along the wire (red triangles). On the coverslip, velocities are distributed isotropically while they populate more densely around the 0 and 180 degree direction of the wire axis, indicating that microtubules move mostly longitudinally along the glass wire. We also examined the dependence of the directional anisotropy of movement on the diameter of the wire. As the wire diameter increases, the velocity distribution becomes more isotropic, which is likely due to the reduction in the surface curvature for the larger diameter. As a result, the gliding behavior approaches the case of the flat coverslip (Figure 5b). When a microtubule binds in a direction transverse to the wire's axis, it needs to bend. The bending energy will compete with the binding energy of the microtubule to the wire. As a result, binding of microtubules along the wire's axis will become more favorable as the wire diameter decreases. The data in Figure 5b is qualitatively consistent with this model, however, the number of bins in this work is too small to make any quantitative conclusions.

Distributions of the gliding speed on the coverslip and on the wire are shown in Figure 5c. The average speed on the glass wire (mean ± sd; 37 ± 11 nm/sec) is reduced by ~13 % compared to the average speed on the coverslip (42 ± 11 nm/sec). The reduction in the gliding speed on the wire may in part be due to the fact that the microtubule gliding motion on the wire was projected on the imaging plane. The average gliding speed on the coverslip is also lower than the previously published value (230 ± 70 nm/sec) that used a ~400-residue *Drosophila* kinesin construct (Brendza et al., 1999). Possible causes for the reduction in speed are a different surface treatment (Ozeki et al., 2009; Bieling et al., 2008), an excessive number of motor proteins (Bieling et al., 2008; Gibbons et al., 2001), a loss of functionality upon surface immobilization or a non-optimal quality of the prepared proteins.

An example, as to how the capability of the glass wire as a transport element enables new functionality is shown in Figure 6, where it (1) bridges two separate systems in a common bath and (2) combines macroscopic fluid injection, i.e. regular pipetting, while allowing molecular-level delivery. In the flow cell used for the demonstration a glass wire is connecting two separate chip surfaces (See the schematic diagram in Figure 6h). Upon carrying out kinesin treatment of the flow cell and introduction of molecular shuttles (microtubules) by regular pipetting, the molecular shuttles trapped on the kinesin-treated glass wire are driven towards the two chip surfaces along the specific paths. Microtubules gliding along the wire bridge are visualized by the images in Figure 6 (a~f, See also Supplemental Movies 13 and 14). This simple demonstration may ensure that the glass wire-based approach is applicable for building three-dimensional fluidic device architectures which require pathways for molecular shuttle-mediated cargo transportation between a main fluidic channel and sub-systems or between sub-systems.

## 4. Conclusions

Kinesin-coated glass wires can efficiently align microtubules along their axis. The glass wire thus serves as a guiding track whose contour can be externally adjusted due to its flexibility. This new approach facilitates bio-micro/nano devices requiring advanced capabilities, such as instantaneous transformation of the track configuration, three dimensional delivery, active aiming, and chip-to-chip communication. With wire diameters down to sub-micrometer range this method is complementary, in terms of the dimension and flexibility, to other methods using nanometer scale wire-type guiding templates (Sikora et al., 2014; Byun et al., 2009; ten Sietoff et al., 2013). The glass wire-based approach presented here is, in general, applicable to different molecular motility systems, e.g. the actomyosin system which also has been extensively investigated for molecular motor-based devices. Characteristics, such as a range of motion speeds and flexural rigidity, of each of the systems are complementary to each other (Agarwal and Hess, 2010; Månsson et al., 2012). Thus the glass wire templates, with its inherent simplicity of production and handling, may be beneficial not only for interfacing macroscale fluidic delivery with molecular-level transportation, but also for potential hybridization of different motility systems. Furthermore, the cylindrical curvature of a glass wire may also allow the study of fundamental phenomena such as the effect of geometrical constraints on microtubule dynamics and assembly, which has relevance to the behaviors in the cell such as in tunneling nanotubes (Gerdes and Carvalho, 2008). Further quantification of the measurements and more detailed analysis with a wider range of wire diameters may yield additional information about the gliding behavior of microtubules on this quasi one-dimensional track.

## Acknowledgements

We gratefully acknowledge support from the World Premier International Research Center Initiative (WPI), MEXT, Japan. We would like to thank Dr. Hideaki Sanada for generously providing the kinesin plasmid.

**Figure captions**

**Fig.1 (a)** Photograph of experimental setup for tapering the glass wire using a flame. **(b)** Schematic diagrams showing the fabrication process of a glass wire. **(c, d)** Fluorescence microscope images of quantum dot decorated glass wires with lateral dimensions 2.56 μm (c) and 0.59 μm (d). The green color-coding for the quantum dots was arbitrarily chosen using Metamorph. **(e)** and **(f)** Intensity profiles along the white dotted lines in the images, c and d, respectively.

**Fig.2 (a)** A bird's eye view of the flow cell including glass wires. **(b)** Schematic cross-sectional view. **(c)** Optical microscope image of the flow cell showing two glass wires crossing over each other.

**Fig.3** Fluorescence images of microtubules in a flow cell including a glass wire. Each image shows the same region (area: ~50 μm × 50 μm). The level of the focal plane is depicted below each image. The white dotted lines in **(a, c)** were added as guidelines for the glass wire based on the image b. **(a)** Microtubules on the coverslip. **(b)** Focal plane at the middle of the glass wire. The light reflected from the glass wire's edge makes it easy to identify the outline. **(c)** Microtubules bound to the top of the glass wire. The cyan color-coding for microtubules was arbitrarily chosen using Metamorph.

**Fig.4** Fluorescence images showing time evolution of the distribution of microtubules in a flow cell (Supplementary Movie 03). White dotted lines denote the glass wire identified in panel b. **(a, h)** The focal plane was set on the coverslip in order to observe microtubules bound on the coverslip. **(b~g)** The focal plane was set on the upper portion of the glass wire to observe microtubules bound on the top. The arrow in panel b indicates the reference vector used to measure the gliding direction of microtubules in Figure 5. Numbers in each image indicate time (min: sec). Arrows in panels **c~g** are guides for the eye for a gliding microtubule.

**Fig.5 (a)** Distribution of microtubule gliding velocities. Blue squares (red triangles) are for microtubules gliding on the coverslip (glass wire). **(b)** Velocity distribution sorted by wire diameter. The orientation of the glass wire is defined relative to the arrow in Figure 4b. **(c)** Histograms of microtubule gliding speeds. The average speed ($\bar{v}$) for the two different cases is indicated in the graph with respective standard deviations ($\sigma$).

**Fig.6** (a~f) Fluorescence and bright-field images showing movement of microtubules bound to a kinesin-coated glass wire connecting two separate chip surfaces (coverslips in panel h). The white dotted lines in panel a denote the edges of the chip surfaces. Numbers in each images indicate time (min:sec). (g) Bright-field image of the glass wire bridging the two chip surfaces. (h) Schematic cross-sectional view of the flow cell.

**Supplementary movies captions**

**SM01**: A glass wire decorated by quantum dots in a buffer solution (Images ~100 μm × 100 μm in size).

**SM02~SM11**: Microtubules gliding on the kinesin coated glass wire. The focus was adjusted occasionally in each movie (Images ~50 μm × 50 μm in size, Actual elapsed time 4 min 32 sec).

**SM12**: Microtubules gliding on the kinesin coated surface of the coverslip (Images ~50 μm × 50 μm in size, Actual elapsed time 4 min 32 sec).

**SM13** and **SM14**: Microtubules gliding towards two separate chip surfaces along the kinesin coated glass wire bridges (Images ~50 μm × 50 μm in size, time tagged on the images (min:sec)).

Figure 1

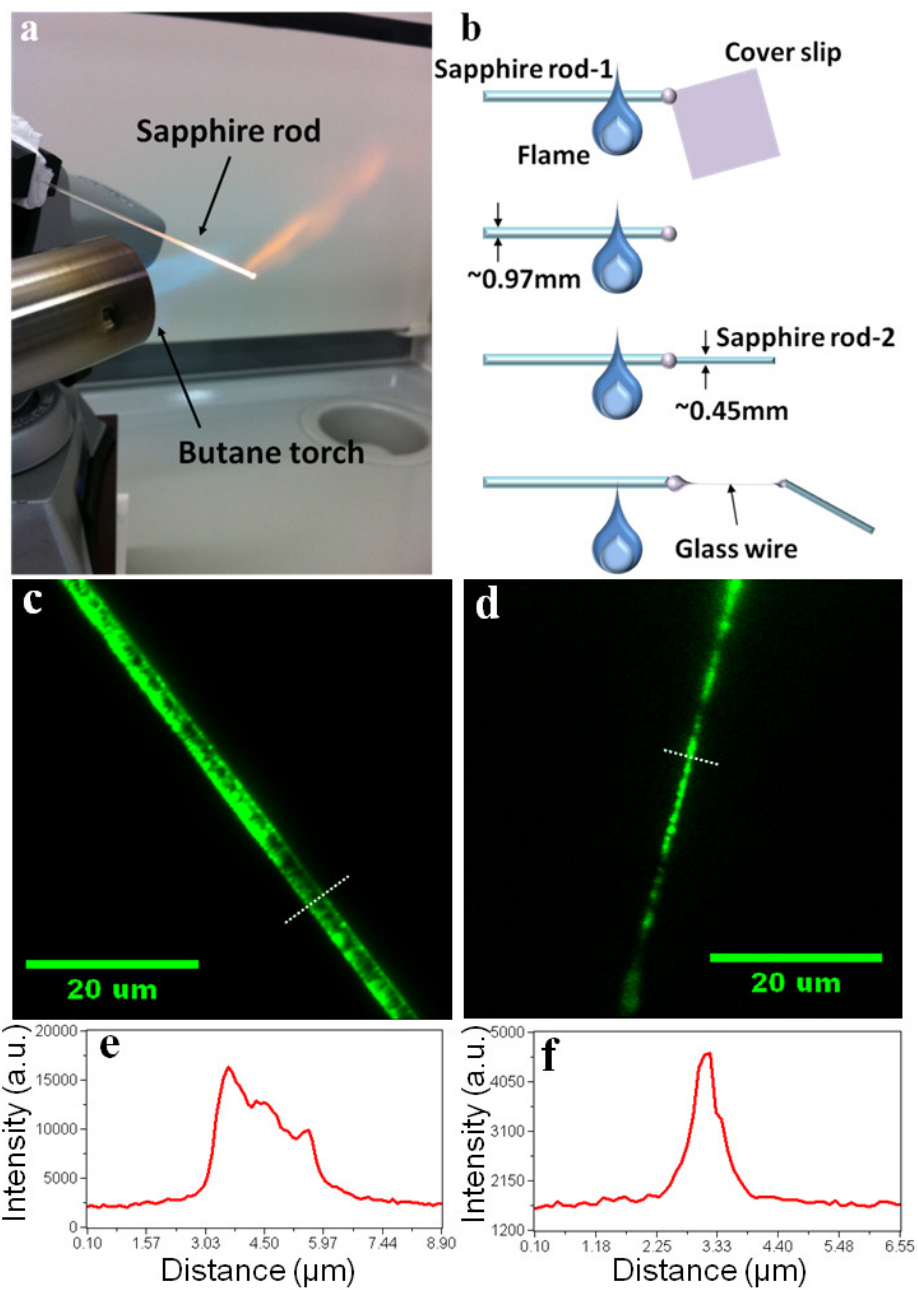

Figure 2

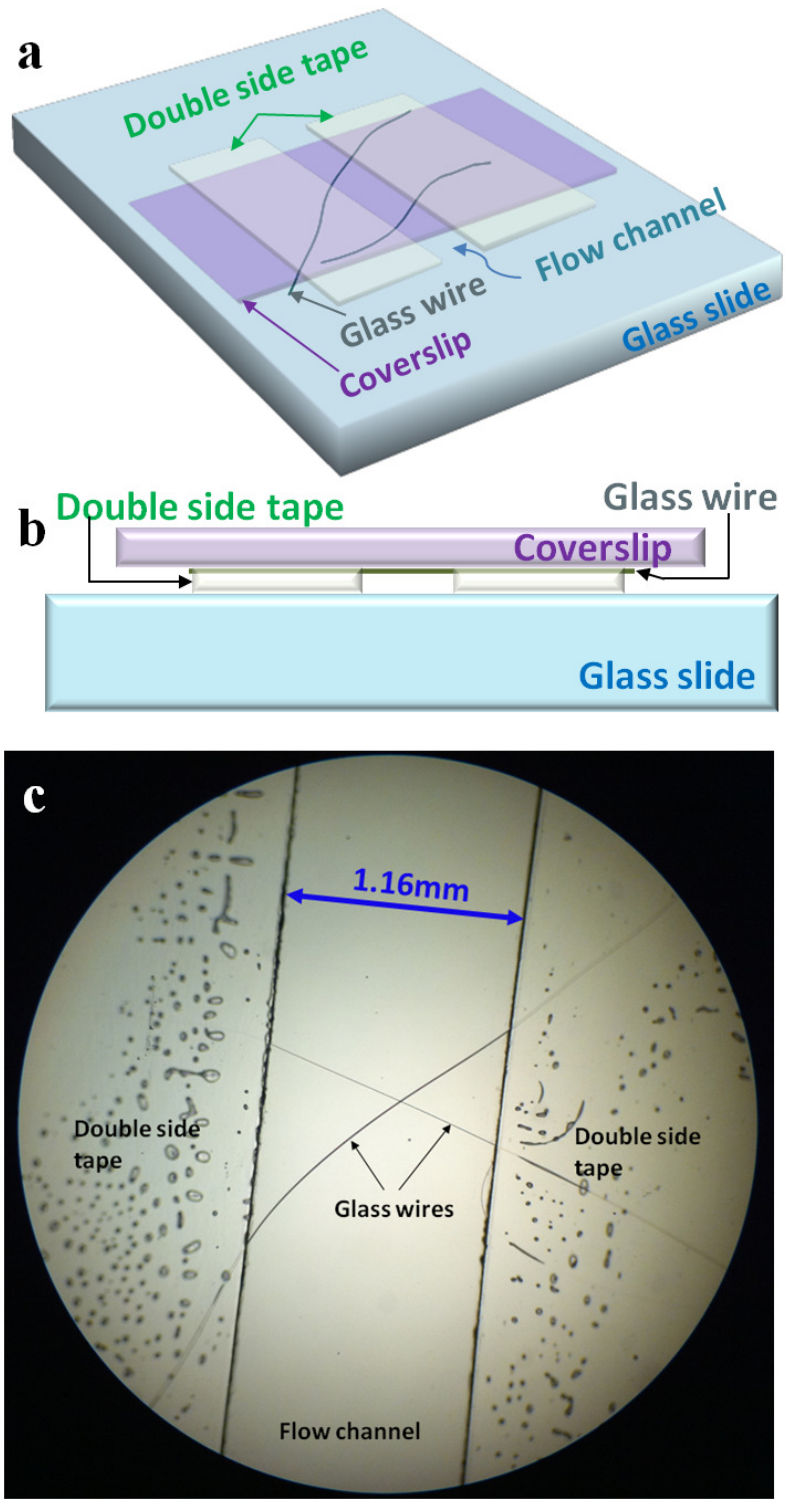

Figure 3

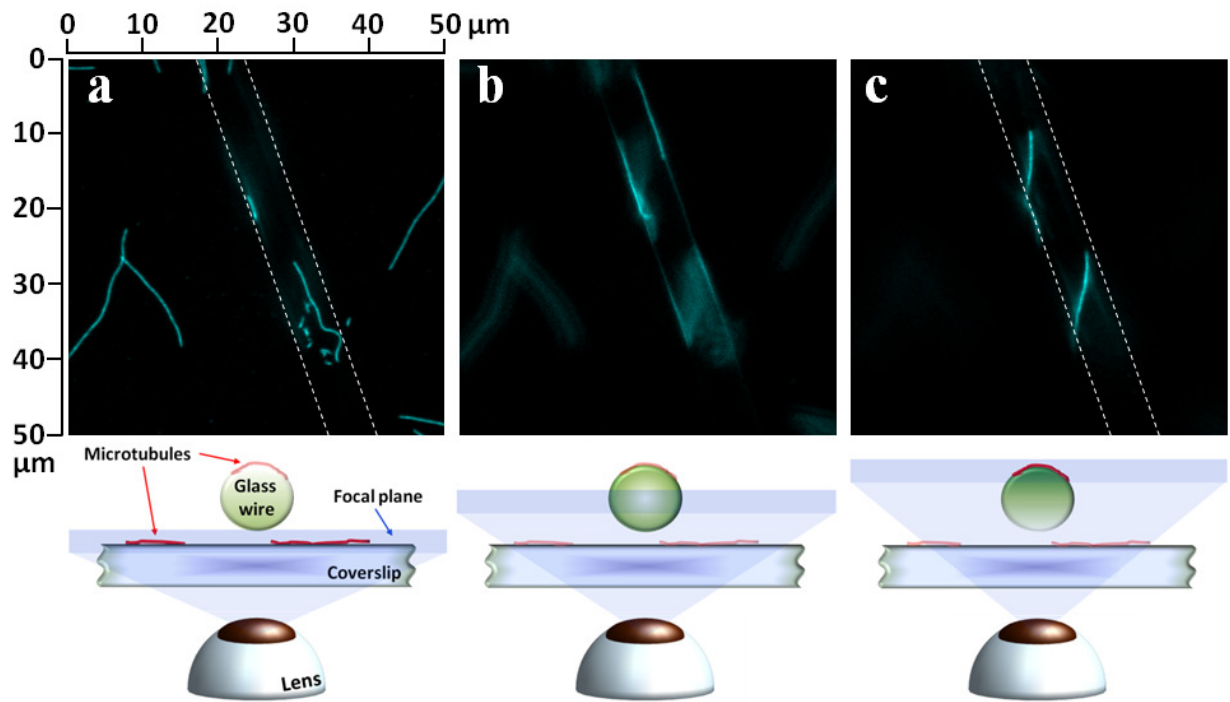

Figure 4

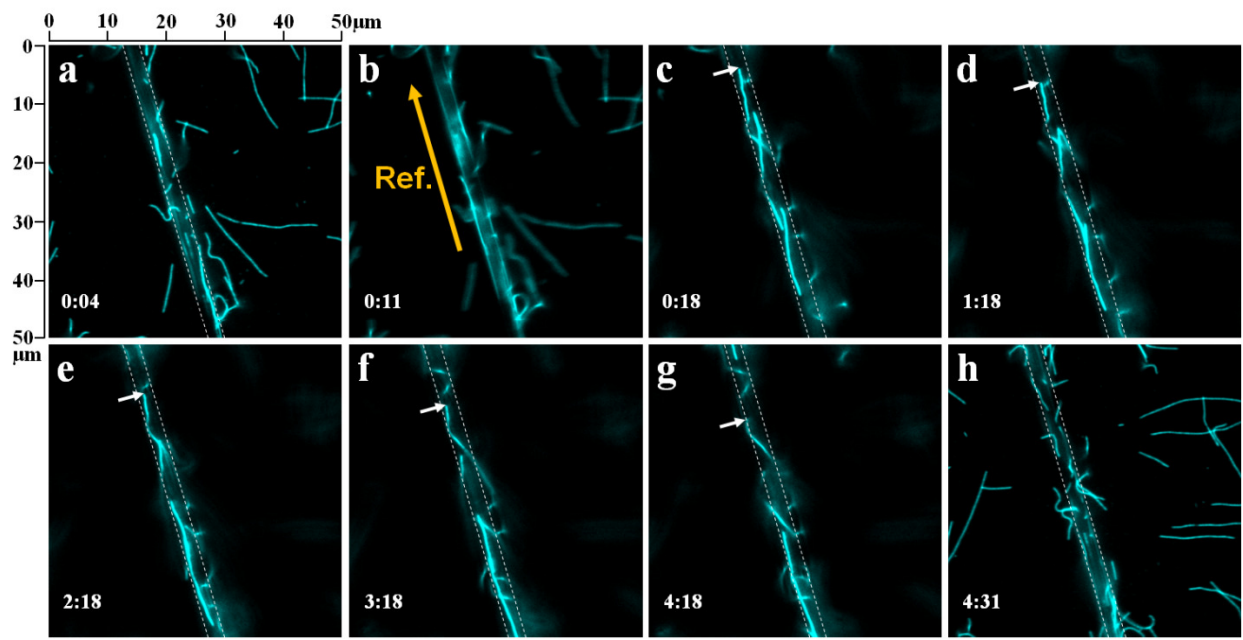

Figure 5

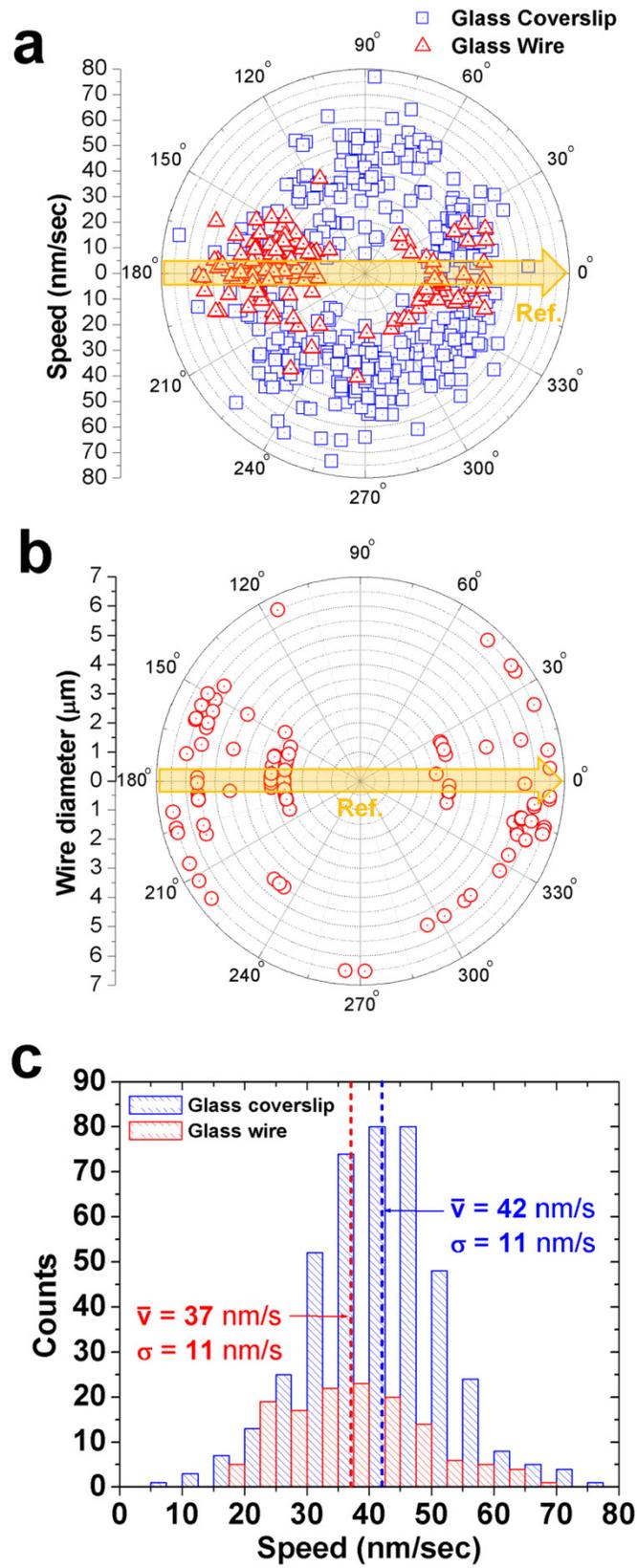

Figure 6

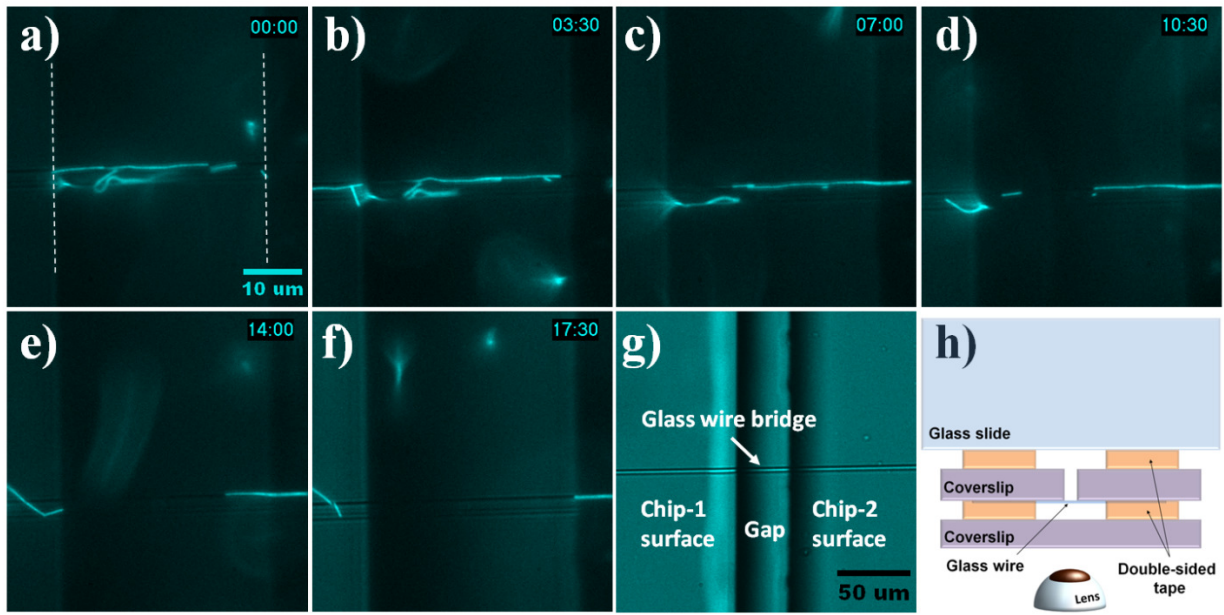